\documentclass[showpacs,floatfix,superscriptaddress]{revtex4}
\usepackage{graphicx}
\usepackage{epsfig}
\usepackage{bm}
\usepackage[utf8]{inputenc}
\usepackage{amssymb}
\usepackage{float}
\usepackage{amsmath}
\usepackage{dcolumn}
\usepackage{latexsym}
\usepackage{subfig}
\usepackage{amsthm}
\usepackage{array}
\usepackage{framed}
\usepackage{cancel}
\usepackage[colorlinks]{hyperref}
\usepackage[usenames,dvipsnames]{color}
\hypersetup{
     breaklinks=true,
    pdfstartview={FitH},    
    colorlinks=true,       
    linkcolor=blue,          
    citecolor=red,        
    filecolor=magenta,      
    urlcolor=blue,           
    anchorcolor=green,      
    linktocpage=true
}

\def\doi{http://doi.org}

\begin{document}

\title{Accretion onto a static spherically symmetric regular MOG dark compact object}

\author{Kourosh Nozari}
\email[]{knozari@umz.ac.ir}
\author{Sara Saghafi}
\email[]{s.saghafi@umz.ac.ir}
\author{Fateme Aliyan}
\email[]{f.alian@stu.umz.ac.ir}

\affiliation{Department of Theoretical Physics, Faculty of Science, University of Mazandaran,\\
P. O. Box 47416-95447, Babolsar, Iran}
\affiliation{ICRANet-Mazandaran, University of Mazandaran, P. O. Box 47416-95447, Babolsar, Iran}

\begin{abstract}
In astrophysics, the process of a massive body acquiring matter is referred to as accretion. The extraction of gravitational energy occurs as a result of the infall. Since it converts gravitational energy into radiation, accretion onto dark compact objects, e.g. black holes, neutron stars, and white dwarfs is an extremely significant process in the astrophysical context. Accretion process is a fruitful way to explore the features of modified gravity (MOG) theories by testing the behavior of their solutions associated with dark compact objects. In this paper, we study the motion of electrically neutral and charged particles moving in around a regular spherically symmetric MOG dark compact object to explore their related innermost stable circular orbit (ISCO) and energy flux. Then, we turn to investigate the accretion of perfect fluid onto the regular spherically symmetric MOG dark compact object. We obtain analytical expressions for four-velocity and proper energy density of the accreting fluid. We see that the MOG parameter increases the ISCO radius of either electrically neutral or charged test particles while it decreases the corresponding energy flux. Moreover, the energy density and the radial component of the four-velocity of the infalling fluid decrease by increasing the MOG parameter near the central source.
\vspace{12 pt}
\end{abstract}


\maketitle

\enlargethispage{\baselineskip}
\tableofcontents

\section{Introduction}\label{intro}

Phenomenologically, dark compact objects are an extensive family of astrophysical objects, which include black holes, neutron stars, white dwarfs, etc. In theoretical point of view, dark compact objects could be predicted in the context of extended gravity theories as well as in scenarios of the beyond the standard model of particle physics \cite{Cardoso:2019rvt}. Recently, observations of LIGO/Virgo proved the existence of binary black holes mergers through detection of gravitational waves \cite{LIGOScientific:2016aoc,LIGOScientific:2016sjg,LIGOScientific:2016dsl,LIGOScientific:2017bnn}, and additionally the Event Horizon Telescope (EHT) revealed the existence of supermassive black holes in center of galaxy M87 \cite{EventHorizonTelescope:2019dse,EventHorizonTelescope:2019uob,EventHorizonTelescope:2019jan,EventHorizonTelescope:2019ths,EventHorizonTelescope:2019pgp,
EventHorizonTelescope:2019ggy,EventHorizonTelescope:2021bee,EventHorizonTelescope:2021srq} and Milky Way \cite{EventHorizonTelescope:2022wkp,EventHorizonTelescope:2022apq,EventHorizonTelescope:2022wok,EventHorizonTelescope:2022exc,EventHorizonTelescope:2022urf,
EventHorizonTelescope:2022xqj}. Therefore, it can be naturally anticipated that future advances in the field of gravitational wave astronomy and very long baseline interferometry will reveal new species of compact objects. On the other hand, it is fascinating to understand how and at what limits a dark compact object tends to be a black hole by increasing its compactness, which makes interesting the study of dark compact objects from a mathematical viewpoint.

General Theory of Relativity (GR) designed by Albert Einstein, besides a lot of achievements in explaining observations and predicting astonishing phenomena, is not yet the complete theory to describe gravitational interaction and corresponding events in the Universe. Reproduction of the rotation curves of nearby galaxies \cite{Sofue:2000jx,Sofue2016}, mass profiles of galaxy clusters \cite{Ettori2013,Voigt2006}, intrinsic singularities at the center of black holes, etc are some examples of the failures of GR. Additionally, GR requires the cosmological constant term $\Lambda$ to explain the positively accelerated expansion of the Universe at late-time \cite{Weinberg1989,Peebles2003}. One interesting way to reform GR is to restructure the geometric part of GR through different approaches that can e.g. result in the so-called MOdified Gravity (MOG), which is a Scalar-Tensor-Vector (STVG) theory to describe gravitational interaction \cite{Moffat2006}, proposed and developed by John W. Moffat. A massive vector field $\phi$ in addition to three scalar field as the mass of the vector field $\tilde{\mu}$, the effective gravitational constant $G$, and the vector field coupling $\xi$ are responsible for expressing the gravitational effects of spacetime in MOG setup. The MOG theory has several achievements in describing astrophysical observations, such as clarifying the rotation curves of many galaxies and the dynamics of galactic clusters without dark matter \cite{Brownstein2006a,Brownstein2006b,Brownstein2007,Moffat2013,Moffat2014,Moffat2015a}, in addition to compatibility with Planck 2018 data \cite{Moffat2021a}. Moreover, several black hole solutions including non-rotating and rotating ones \cite{Moffat2015b} even with extra dimensions \cite{Cai2021}, cosmological solutions \cite{Roshan2015,Jamali2018,Davari2021} and also, non-stationary solutions for inhomogeneity distributions of mass-energy in spacetime \cite{Perez2019} are released within the framework of MOG theory in recent years. Also, many theoretical and observational efforts have been done to understand the MOG theory features and how it work in different situations \cite{Moffat2021b,Guo2018,Moffat2020,Mureika2016,Saghafi2021,Saghafi:2021wzx,Perez:2017spz,Shukla:2022sti,Hussain:2015cga,Sharif:2017owq,Lee:2017fbq,Kolos:2020ykz,
Hu:2022lek,John:2019was}. Interestingly, the solution describing the regular rotating and non-rotating MOG dark compact object has been recently explored in Ref. \cite{Moffat:2018jmi}. The shadow behaviour of the regular rotating and non-rotating MOG dark compact object is investigated in Ref. \cite{Sau:2022afl}.

Accretion is a process of particles being dragged onto a dark compact object. This process releases extra energy into surroundings, which is a source of some astronomical phenomena \cite{Kato:20008,Martnez:2014}; for instance the production of powerful jets, high-energy radiation, and quasars. A flattened structure made by rotating gaseous materials that slowly spiral into a massive central body is called an accretion disk. Accretion disks typically form around compact objects when interstellar matter exists. Accretion disks of compact objects are results of rotating gaseous materials in unstable bounded orbits \cite{Kato:20008,Martnez:2014}. Under some conditions, the gas particles fall into gravitational potential of the compact objects,  which causes gravitational energy in the form of heat. The inner portion of the accretion disk cools down as a result of the conversion of some heat into radiation \cite{Kato:20008,Martnez:2014}. The electromagnetic spectrum of the emitted radiation can be analyzed when it reaches radio, optical, or X-ray telescopes. The motion of the gas particles, which may also be related to the structure and nature of the central mass, determines the properties of this radiation. As a result, studying accretion disk emission spectra can provide fruitful astrophysical data. Hence the accretion disks of compact objects drawn a lot of attention and have been studied in several cases in the literature \cite{Michel:1972,Jamil:2008bc,Babichev:2008dy,JimenezMadrid:2005rk,Babichev:2008jb,Giddings:2008gr,Sharif:2011ih,John:2013bqa,Debnath:2015yva,Ganguly:2014cqa,Mach:2013fsa,
Mach:2013gia,Karkowski:2012vt,Yang:2015sfa,Babichev:2004yx,Babichev:2005py,Gao:2008jv,Barausse:2018vdb,Nozari:2020swx,Zheng:2019mem,Salahshoor:2018plr,Jiao:2016iwp}.

The regular non-rotating (spherically symmetric) MOG dark compact object \cite{Moffat:2018jmi}, which can be formed from the collapse of stellar object, can be tested in astrophysical phenomena, e.g. accretion process. It is the reason we found it interesting to study accretion disk onto the regular MOG dark compact object. In this regard, we also aim to study the motion of electrically neutral and charged test particles moving in this spacetime, and explore their corresponding energy flux. The rest of the paper is organized as follows. In Section \ref{STVG}, we review the MOG field equations, and then, we introduce the regular MOG dark compact object spacetime and its features. Next, we study the motion of electrically neutral and charged test particles travelling in the regular MOG dark compact object spacetime in Section \ref{MTP}. Then, we investigate the static spherically symmetric accretion in Section \ref{ARMDCO}. Finally, we end with some conclusions in Section \ref{SaC}.

\section{Action and field equations of STVG theory}\label{STVG}

The total action in the theory of STVG is in the form of \cite{Moffat2006}
\begin{equation}\label{tact}
S=S_{GR}+S_{M}+S_{\phi}+S_{S}\,,
\end{equation}
where $S_{GR}$ is the Einstein-Hilbert action, $S_{M}$ is the action of all possible matter sources, $S_{\phi}$ is the action of the (spin $1$ graviton) vector field $\phi^{\mu}$ possessing the mass $\tilde{\mu}$ as one of the scalar fields in the theory, and $S_{S}$ is the action of three scalar fields, which can be expressed as follows
\begin{equation}\label{sgr}
S_{GR}=\frac{1}{16\pi}\int d^{4}x\sqrt{-g}\frac{1}{G}R\,,
\end{equation}
\begin{equation}\label{sphi}
S_{\phi}=-\int d^{4}x\sqrt{-g}\left(\frac{1}{4}B^{\mu\nu}B_{\mu\nu}+V_{1}(\phi)\right)\xi\,,
\end{equation}
\begin{equation}\label{ss}
\begin{split}
S_{S} & =\int d^{4}x\sqrt{-g}\left[\frac{1}{G^{3}}\left(\frac{1}{2}g^{\mu\nu}\nabla_{\mu}G\nabla_{\nu}G-V_{2}(G)\right)
+\frac{1}{\tilde{\mu}^{2}G}\left(\frac{1}{2}g^{\mu\nu}\nabla_{\mu}\tilde{\mu}\nabla_{\nu}\tilde{\mu}-V_{3}(\tilde{\mu})\right)\right.\\
& \left.+\frac{1}{G}\left(\frac{1}{2}g^{\mu\nu}\nabla_{\mu}\xi\nabla_{\nu}\xi-V_{4}(\xi)\right)\right]\,,
\end{split}
\end{equation}
in which $g_{\mu\nu}$ is the background metric tensor and $g$ is the corresponding determinant, $R$ is the Ricci scalar constructed by contracting $R_{\mu\nu}$ as the Ricci tensor, $G$ is a scalar field in the setup, which known as the enhanced Newtonian parameter, $\xi$ is third scalar field in the setup as the vector field coupling, $V_{1}(\phi)$, $V_{2}(G)$, $V_{3}(\tilde{\mu})$, and $V_{4}(\xi)$ are the corresponding potentials of the vector field $\phi^{\mu}$, and three scalar field $G$, $\tilde{\mu}$, and $\xi$, respectively, and $B_{\mu\nu}=\partial_{\mu}\phi_{\nu}-\partial_{\nu}\phi_{\mu}$, and also $\nabla_{\mu}$ stands for the covariant derivative in the spacetime.

In the STVG theory, $T_{\mu\nu}={}^{(M)}T_{\mu\nu}+{}^{(\phi)}T_{\mu\nu}+{}^{(S)}T_{\mu\nu}$ is the total stress-energy tensor, in which the stress-energy tensor of matter sources is ${}^{(M)}T_{\mu\nu}$, the stress-energy tensor of the scalar fields is ${}^{(S)}T_{\mu\nu}$, and the stress-energy tensor of the vector field is
\begin{equation}\label{setphi}
{}^{(\phi)}T_{\mu\nu}=-\frac{1}{4}\left(B_{\mu}^{\,\,\,\sigma}B_{\nu\sigma}-\frac{1}{4}g_{\mu\nu}B^{\sigma\lambda}B_{\sigma\lambda}\right)\,,
\end{equation}
for which $V_{1}(\phi)=0$. One can find the full field equations of the STVG framework by variation of the action $S$ concerning the inverse of the metric tensor, which yields \cite{Moffat2006}
\begin{equation}\label{ffe}
G_{\mu\nu}+G\left(\nabla^{\gamma}\nabla_{\gamma}\frac{1}{G}g_{\mu\nu}-\nabla_{\mu}\nabla_{\nu}\frac{1}{G}\right)=8\pi G T_{\mu\nu}\,,
\end{equation}
in which the Einstein tensor is defied as $G_{\mu\nu}=R_{\mu\nu}-\frac{1}{2}g_{\mu\nu}R$.

\subsection{Regular MOG static spherically symmetric dark compact object}

The line element of the regular MOG static spherically symmetric dark compact object were found under the following assumptions \cite{Moffat:2018jmi}
\begin{itemize}
  \item The vector field is massless, i.e., $\tilde{\mu}=0$, since one can prove that for MOG compact objects, e.g., black holes possessing horizons, the mass of the vector field in the setup is zero.
  \item The enhanced Newtonian parameter $G$ is defined as a constant depending on the free dimensionless parameter $\alpha$ so that $G=G_{N}(1+\alpha)$ where $G_{N}$ is the Newtonian constant. Furthermore, the gravitational source charge of the vector field is $Q_{g}=\sqrt{\alpha G_{N}}M$ where $M$ is the source mass. Here, we set $G_{N}=1$.
  \item The vector field coupling is set to unity, i.e., $\xi=1$.
  \item The matter-free field equations of STVG setup is considered since the MOG dark compact object is a vacuum solution of the framework.
\end{itemize}
The above assumptions result in $S_{M}=S_{S}=0$ and consequently, we have ${}^{(M)}T_{\mu\nu}={}^{(S)}T_{\mu\nu}=0$. Thus, the field equations \eqref{ffe} now reduce to the following form
\begin{equation}\label{rfe}
G_{\mu\nu}=8\pi(1+\alpha){}^{(\phi)}T_{\mu\nu}\,.
\end{equation}
Solving the last equation by following the procedure introduced in Ref. \cite{Moffat:2018jmi} leads to the line element of the regular MOG static spherically symmetric dark compact object as follows
\begin{equation}\label{le}
ds^{2}=f(r)dt^{2}-\frac{1}{f(r)}dr^{2}-r^{2}d\Omega^{2}\,,
\end{equation}
where $d\Omega^{2}=d\theta^{2}+\sin^{2}\theta d\varphi^{2}$ is the line element of the unit 2-sphere, and also we have defined
\begin{equation}\label{fle}
f(r)=1-\frac{2(1+\alpha)Mr^{2}}{\left(r^{2}+\alpha(1+\alpha)M^{2}\right)^{\frac{3}{2}}}+\frac{\alpha(1+\alpha)M^{2}r^{2}}{\left(r^{2}+\alpha(1+\alpha)M^{2}\right)^{2}}\,,
\end{equation}
which satisfies the weak energy condition \cite{Moffat:2015kva,Ayon-Beato:1998hmi}. The MOG dark compact object possesses a critical value for $\alpha$ as $\alpha_{crit}=0.674$ \cite{Moffat:2018jmi}, so that for $\alpha\leq\alpha_{crit}$ it has two horizons. It is worth mentioning that the (spin $1$ graviton) vector field produces a repulsive gravitational force, which prevents the collapse of the MOG dark compact object to a MOG black hole with horizon.

Setting $\alpha=0$ in the line element \eqref{le} recovers the Schwarzschild black hole in GR. Moreover, the asymptotic behavior of the MOG compact object in the limit of  $r\rightarrow\infty$ is deduced as follows
\begin{equation}\label{fleabl}
f(r)\approx 1-\frac{2(1+\alpha)M}{r}+\frac{\alpha(1+\alpha)M^{2}}{r^{2}}\,.
\end{equation}
When, $\alpha\leq\alpha_{crit}$, the two horizons of the regular MOG static spherically symmetric dark compact object in the limit of $r\rightarrow\infty$ can be found as
\begin{equation}\label{hrz}
r_{\pm}=M\left(1+\alpha\pm\sqrt{1+\alpha}\right)\,.
\end{equation}
When, $\alpha>\alpha_{crit}$, there is a naked regular MOG static spherically symmetric dark compact object with no horizon. On the other hand, approaching the source, i.e., $r\rightarrow 0$, the MOG dark compact object behaves to the form
\begin{equation}\label{fleabs}
f(r)\approx 1-\frac{r^{2}}{M^{2}}\left(\frac{2\sqrt{1+\alpha}-\sqrt{\alpha}}{(1+\alpha)\alpha^{\frac{3}{2}}}\right)\,.
\end{equation}
Therefore, the spacetime metric of the MOG dark compact object is regular so that $f(0)=1$. Additionally, one can verify that the Kretschmann scalar $R^{\mu\nu\lambda\sigma}R_{\mu\nu\lambda\sigma}$ in addition to the Ricci scalar $R$ in the spacetime metric are regular at $r=0$.

For the static spherically symmetric system, the gravitational redshift $z$ at the asymptotic distance $r$ to an observer is gathered as follows
\begin{equation}\label{grs}
z(r)=\frac{1}{\sqrt{f(R)}}-1\,,
\end{equation}
where the radius of the MOG dark compact object is $R$. For $\alpha<\alpha_{crit}$, in the limit of $r\rightarrow\infty$, the gravitational redshift of the compact object becomes infinite on the horizon $r_{+}$ and for $\alpha>\alpha_{crit}$ it has a finite value. Based on the observational data, however, one anticipates that the regular MOG dark compact object is adequately dark to be compatible with binary X-ray observations, so that $\alpha\sim \alpha_{crit}$ \cite{Moffat:2018jmi}.

\section{Motion of test particle in MOG dark compact object spacetime}\label{MTP}

The geodesic structure of the spacetime of the MOG compact object governs the trajectory of a test particle. In this section, we investigate the time-like geodesics around the regular MOG static spherically symmetric dark compact object through Lagrangian formalism \cite{Salahshoor:2018plr,Hussain:2015cga,Shukla:2022sti,Nozari:2020tks,Saghafi:2022pme}.

Under temporal translation and rotation around the axes of symmetry, the line element \eqref{le} of the MOG dark compact object associated with the metric coefficient \eqref{fle} is invariant since this spacetime is static and spherically symmetric. Therefore, the spacetime of the regular MOG dark compact object possesses two Killing vectors as follows
\begin{equation}\label{killvec}
\begin{split}
& {}^{(t)}\zeta^{\mu}\frac{\partial}{\partial x^{\mu}}=(1,0,0,0)\frac{\partial}{\partial x^{\mu}}=\frac{\partial}{\partial t}\,,\\
& {}^{(\varphi)}\zeta^{\mu}\frac{\partial}{\partial x^{\mu}}=(0,0,0,1)\frac{\partial}{\partial x^{\mu}}=\frac{\partial}{\partial\varphi}\,.\\
\end{split}
\end{equation}
These Killing vectors imply two conserved (constants) quantities for the motion of the test particle in the spacetime, which we aim to find them in the following. We plan to investigate the trajectory of both electrically neutral and charged test particles motion around the regular MOG dark compact object.

\subsection{Motion of electrically neutral test particle}

\begin{figure}[htb]
\centering
\includegraphics[width=0.7\textwidth]{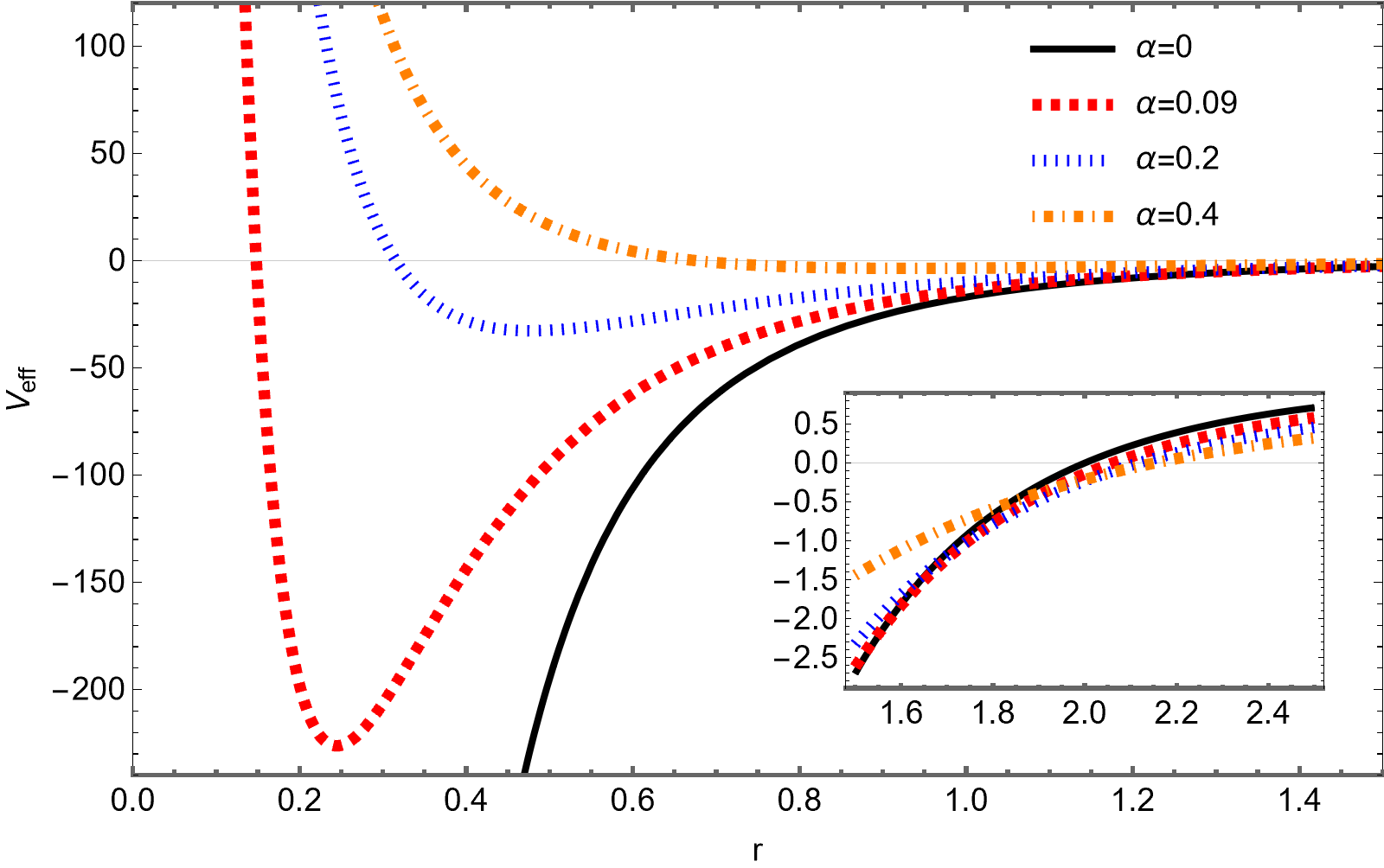}
\caption{\label{Fig1}\small{\emph{The illustration of $V_{eff}$ of the regular static spherically symmetric MOG dark compact object versus $r$ for different values of $\alpha$. The black solid line is for the case of Schwarzschild solution in GR.}}}
\end{figure}

The Lagrangian of a test particle moving in the spacetime of the regular MOG dark compact object is expressed as
\begin{equation}\label{lag}
\mathcal{L}=\frac{1}{2}g_{\mu\nu}\dot{x}^{\mu}\dot{x}^{\nu}\,,
\end{equation}
where over-dot stands for derivative with respect to the affine parameter $\tau$. The four-velocity of the test particle is defined as $\dot{x}^{\mu}\equiv u^{\mu}=(u^{t},u^{r},u^{\theta},u^{\varphi})$. We interested in the planar motion of the particle on the equatorial plane with $\theta=\frac{\pi}{2}$. Thus, utilizing the Euler-Lagrange equation
\begin{equation}\label{eullag}
\frac{d}{d\tau}\left(\frac{\partial\mathcal{L}}{\partial\dot{x}^{\mu}}\right)-\frac{\partial\mathcal{L}}{\partial x^{\mu}}=0\,,
\end{equation}
one can find two conserved quantities of the particle motion corresponding with two Killing vectors as follows
\begin{equation}\label{econs}
\frac{dt}{d\tau}=\dot{t}\equiv u^{t}=\frac{E}{f(r)}=\frac{E}{\left(1-\frac{2(1+\alpha)Mr^{2}}{\left(r^{2}+\alpha(1+\alpha)M^{2}\right)^{\frac{3}{2}}}+\frac{\alpha(1+\alpha)
M^{2}r^{2}}{\left(r^{2}+\alpha(1+\alpha)M^{2}\right)^{2}}\right)}\,,
\end{equation}
\begin{equation}\label{lcons}
\frac{d\varphi}{d\tau}=\dot{\varphi}\equiv u^{\varphi}=\frac{L}{r^{2}}\,,
\end{equation}
where $E$ and $L$ as two conserved quantities are the total energy and the total angular momentum per unit mass of the particle, respectively. Moreover, using the Euler-Lagrange equation, we can find $\frac{d\theta}{d\tau}=\dot{\theta}\equiv u^{\theta}=0$ in addition to
\begin{equation}\label{rcompvel}
\frac{dr}{d\tau}=\dot{r}\equiv u^{r}=\left[-f(r)\left(1-\frac{E^{2}}{f(r)}+\frac{L^{2}}{r^{2}}\right)\right]^{\frac{1}{2}}\,.
\end{equation}
Based on the normalization condition for the four-velocity of the test particle, i.e., $u^{\mu}u_{\mu}=1$ and utilizing Eqs. \eqref{fle}, \eqref{econs}, and \eqref{lcons} one can find
\begin{equation}\label{rdot}
\dot{r}^{2}=E^{2}-V_{eff}\,,
\end{equation}
where $V_{eff}$ is the effective potential of the test particle, which is defined as
\begin{equation}\label{veff}
\begin{split}
V_{eff}=f(r)\left(1+\frac{L^{2}}{r^{2}}\right)=\left(1-\frac{2(1+\alpha)Mr^{2}}{\left(r^{2}+\alpha(1+\alpha)M^{2}\right)^{\frac{3}{2}}}+\frac{\alpha(1+\alpha)
M^{2}r^{2}}{\left(r^{2}+\alpha(1+\alpha)M^{2}\right)^{2}}\right)\left(1+\frac{L^{2}}{r^{2}}\right)\,.
\end{split}
\end{equation}
Effective potential analysis is significant in studying geodesic structure. The location of the circular orbits, for example, is determined by the local extremum of the effective potential. Figure \ref{Fig1} illustrates the behavior of the effective potential $V_{eff}$ for the MOG dark compact object in comparison with the Schwarzschild case in GR. From Fig. \ref{Fig1} we see that increasing the value of the parameter $\alpha$ leads to increment of the effective potential.

\subsubsection{Stable circular orbits around regular MOG dark compact object}

The main characteristic of circular orbits is $\dot{r}=\ddot{r}=0$ or equivalently $u^{r}=\dot{u}^{r}=0$. Hence, from Eqs. \eqref{econs}-\eqref{rcompvel} one can verify that for circular orbits, $E^{2}=V_{eff}$ and consequently, $\frac{dV_{eff}}{dr}=0$ must be satisfied. Solving these two equations simultaneously by using Eqs. \eqref{fle}, \eqref{econs}, and \eqref{lcons} results in the following relations for the total (specific) energy $E$, total (specific) angular momentum $L$, and the angular velocity $\Omega_{\varphi}\equiv\frac{d\varphi}{dt}=\frac{u^{\varphi}}{u^{t}}$ for the test particle in MOG dark compact object background
\begin{equation}\label{E2}
\begin{split}
E^{2} & =\frac{2f^{2}(r)}{2f(r)-rf'(r)}=\frac{\left(\alpha(1+\alpha)M^{2}+r^{2}\right)^{3}}{y_{1}}\left(1-\frac{2(1+\alpha)Mr^{2}}{\left(r^{2}+\alpha(1+\alpha)M^{2}\right)
^{\frac{3}{2}}}+\frac{\alpha(1+\alpha)M^{2}r^{2}}{\left(r^{2}+\alpha(1+\alpha)M^{2}\right)^{2}}\right)^{2}\,,
\end{split}
\end{equation}
\begin{equation}\label{L2}
L^{2}=\frac{r^{3}f'(r)}{2f(r)-rf'(r)}=\frac{y_{2}(1+\alpha)Mr^{4}}{y_{1}\sqrt{\alpha(1+\alpha)M^{2}+r^{2}}}\,,
\end{equation}
\begin{equation}\label{Omega2}
\Omega_{\varphi}^{2}=\frac{1}{2r}f'(r)=\frac{y_{2}(1+\alpha)M}{\left(\alpha(1+\alpha)M^{2}+r^{2}\right)^{\frac{7}{2}}}\,,
\end{equation}
where a prime stands for differentiation with respect to radial coordinate $r$ and also we have defined
\begin{equation}\label{y1}
y_{1}\equiv r^{6}+\alpha^{3}(1+\alpha)^{3}M^{6}+3\alpha^{2}(1+\alpha)^{2}M^{4}r^{2}+(1+\alpha)Mr^{4}\left(5\alpha M-3\sqrt{\alpha(1+\alpha)M^{2}+r^{2}}\right)\,,
\end{equation}
\begin{equation}\label{y2}
y_{2}\equiv r^{4}-\alpha Mr^{2}\left(y_{1}+(1+\alpha)M\right)-\alpha^{2}(1+\alpha)M^{3}\left(2(1+\alpha)M-y_{1}\right)\,.
\end{equation}
According to Eqs. \eqref{E2}-\eqref{Omega2} one can see that the condition $2f(r)-rf'(r)>0$ for existence of the circular orbits is required in order the total energy and the total angular momentum to be real.

\begin{figure}[htb]
\centering
\includegraphics[width=0.7\textwidth]{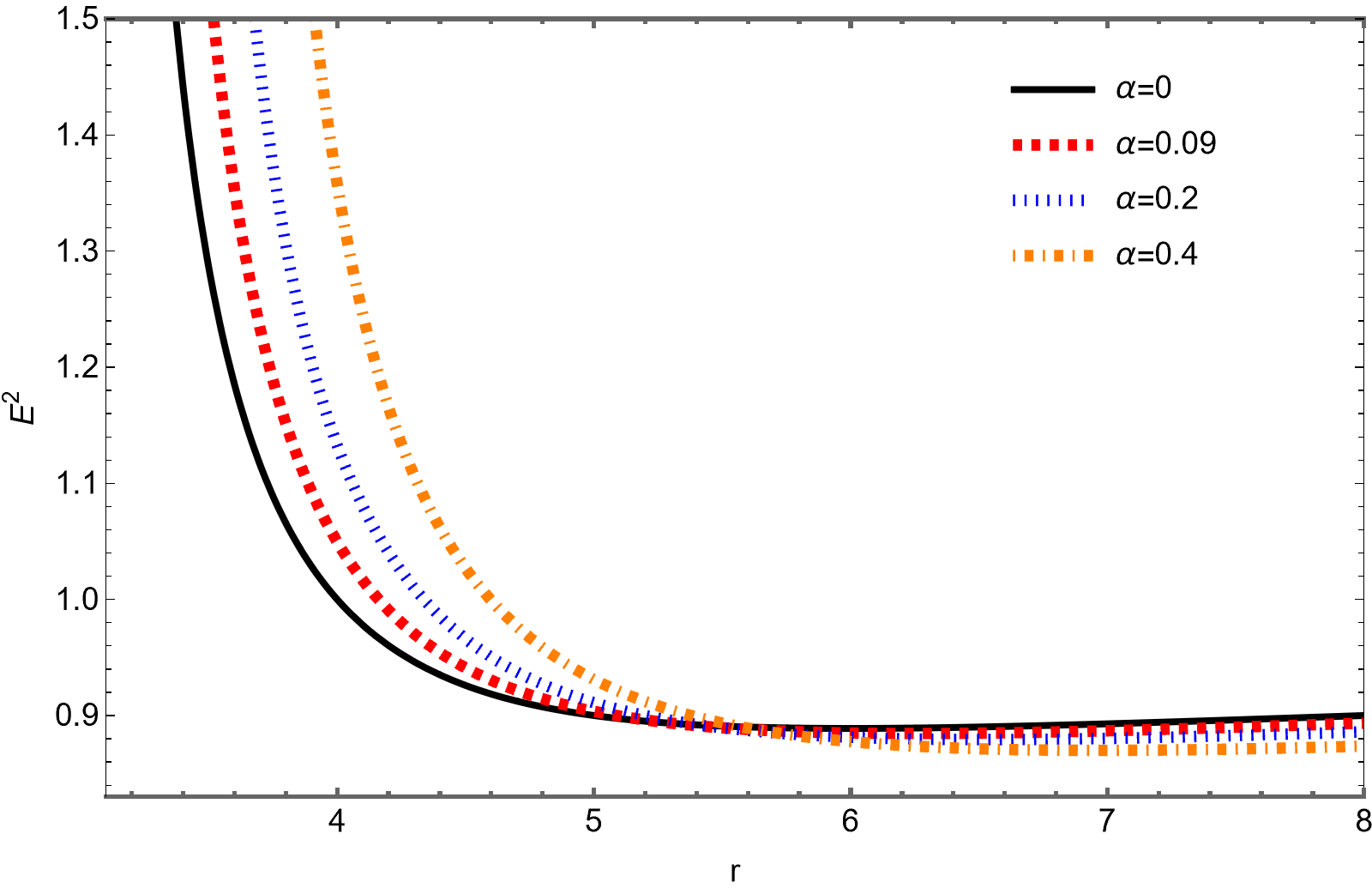}
\caption{\label{Fig2}\small{\emph{The plot of $E^{2}$ of the regular static spherically symmetric MOG dark compact object versus $r$ for different values of $\alpha$. The black line is for the case of Schwarzschild solution in GR.}}}
\end{figure}

\begin{figure}[htb]
\centering
\includegraphics[width=0.7\textwidth]{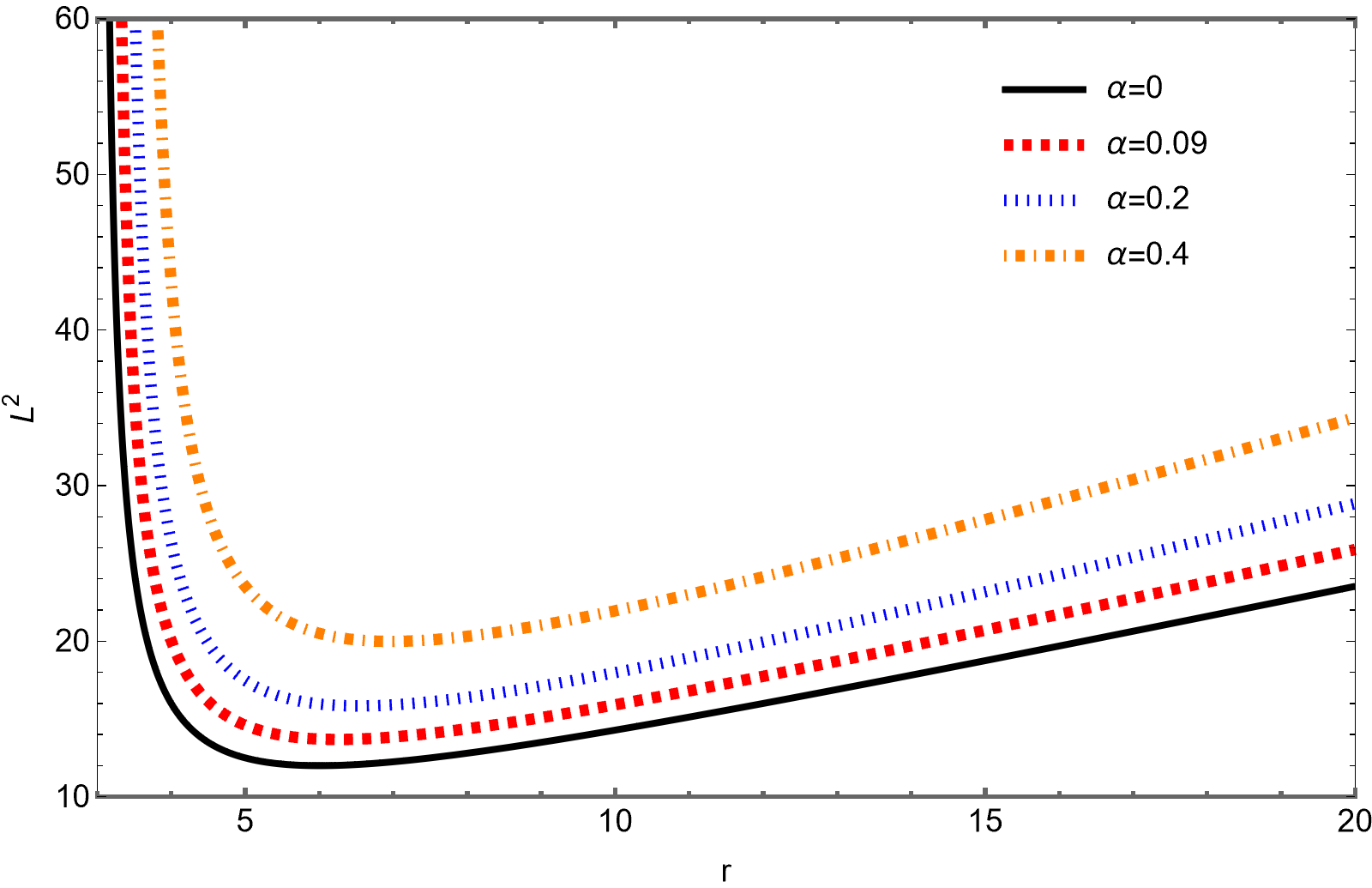}
\caption{\label{Fig3}\small{\emph{The behavior of $L^{2}$ of the regular static spherically symmetric MOG dark compact object versus $r$ for different values of $\alpha$. The black solid line is for the case of Schwarzschild solution in GR.}}}
\end{figure}

\begin{figure}[htb]
\centering
\includegraphics[width=0.7\textwidth]{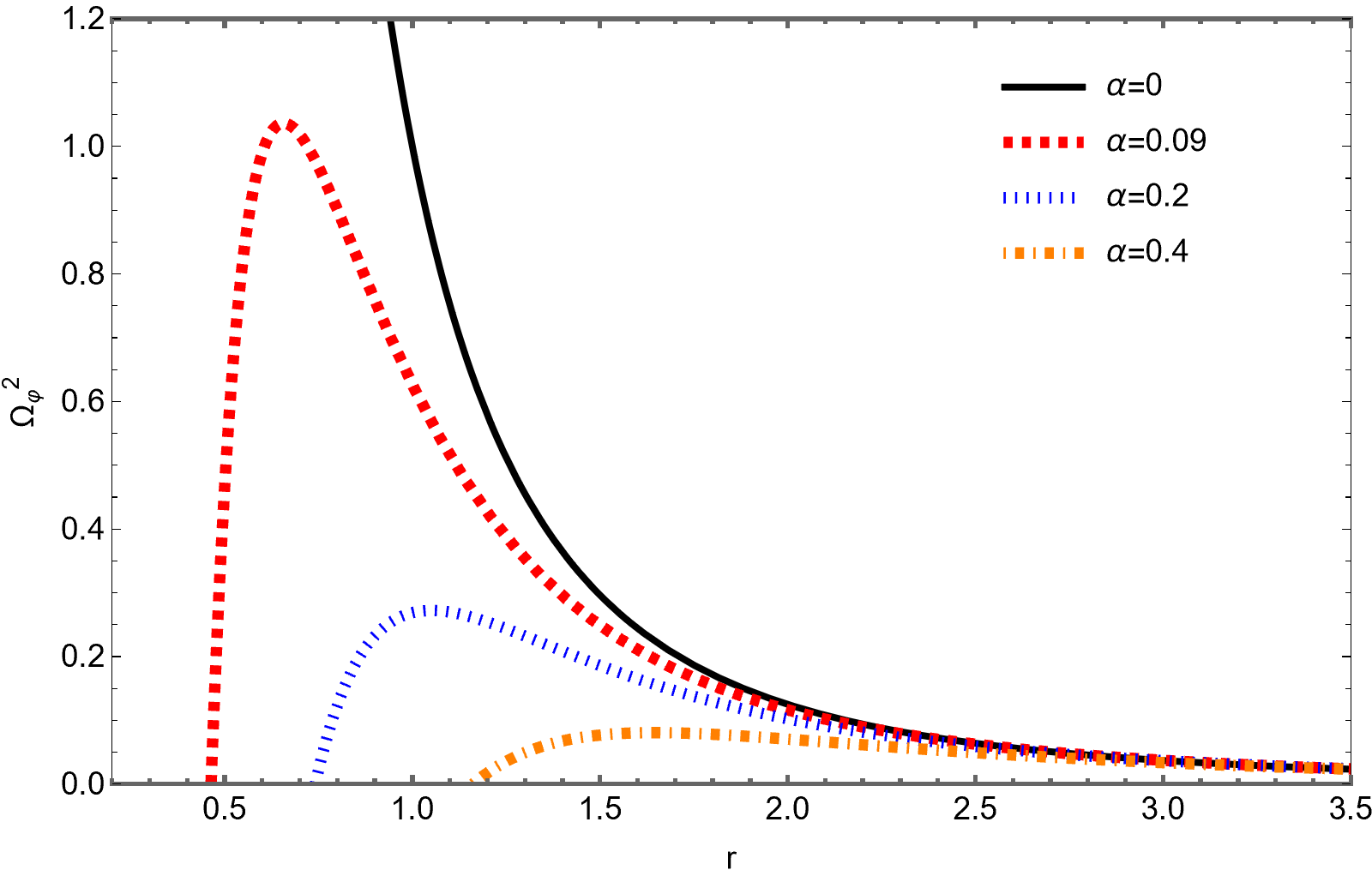}
\caption{\label{Fig4}\small{\emph{The illustration of $\Omega_{\varphi}^{2}$ of the regular static spherically symmetric MOG dark compact object versus $r$ for different values of $\alpha$. The black solid line is for the case of Schwarzschild solution in GR.}}}
\end{figure}

Figure \ref{Fig2} demonstrates the behavior of the $E^{2}$ versus $r$ from which we can see that growing the parameter $\alpha$ leads to amplify the specific energy of the test particle in the spacetime of the regular MOG dark compact object while far from the source, the energy becomes almost constant. The corresponding curve of the Schwarzschild solution in GR is also shown in Fig. \ref{Fig2} which has always smaller values than the regular MOG dark compact object case.
Figure \ref{Fig3} is the illustration of $L^{2}$ versus $r$ associated with the regular MOG dark compact object in comparison with Schwarzschild solution in GR for different values of $\alpha$, so that again increasing it results in growing the value of $L^{2}$. All of these figures have smaller values of $E^{2}$ and $L^{2}$ than the corresponding ones in the case of the Schwarzschild solution in GR.
In Fig. \ref{Fig4} we see the curves of $\Omega_{\varphi}^{2}$ versus $r$ for the regular MOG dark compact object in comparison with the Schwarzschild case in which we see that increasing $\alpha$ leads to reduction of the value of $\Omega_{\varphi}^{2}$ so that the curve of the Schwarzschild case contains higher values of $\Omega_{\varphi}^{2}$ than corresponding ones in the regular MOG dark compact object.

The location of the stable circular orbits correspond to the local minimum of the effective potential. Accordingly, an innermost (marginally) stable circular orbit (ISCO) needs the conditions
\begin{equation}\label{iscocons}
\frac{dV_{eff}}{dr}=0\,,\qquad \frac{d^{2}V_{eff}}{dr^{2}}=0\,,
\end{equation}
to be satisfied. The existence of ISCO, $r_{_{ISCO}}$ is purely a relativistic phenomenon. Instead of classical mechanics in which the effective potential possesses just one minimum; in GR however, the effective potential can generate either a local maximum and minimum or no extremum, relying on the choice of $L$ in the effective potential. A stable outer and an unstable inner circular orbit for the test particle is related to this extremum. ISCO is where the stable and unstable circular orbits coincide for a specific value of $L$. Due to the complexity of the metric coefficient function \eqref{fle} the explicit analytical form of ISCO associated with the regular MOG dark compact object is not available. Hence, solving equation set \eqref{iscocons} numerically by using Wolfram Mathematica (v13.1) results in numerical values of the ISCO for the test particle moving in the spacetime of the MOG dark compact object. To do this, we set $M=1$. Then, for three different values of the MOG parameter $\alpha$ in Table \ref{Table1} we collect the numerical values of $r_{_{ISCO}}$, $L_{_{ISCO}}$, and $E_{_{ISCO}}$ for the regular static spherically symmetric MOG dark compact object. On the other hand, we know that for Schwarzschild black hole in GR, the ISCO is $r_{_{ISCO}}=6M$. Therefore, from Table \ref{Table1} we see that increasing the value of $\alpha$ leads to grow the ISCO associated with the regular MOG dark compact object.
\begin{table}[htb]
  \centering
  \caption{\label{Table1}\small{\emph{The numerical values of $r_{_{ISCO}}$, $L_{_{ISCO}}$, and $E_{_{ISCO}}$ for a test particle moving in the regular static spherically symmetric MOG dark compact object spacetime associated with different values of $\alpha$.}}}
  {\renewcommand{\arraystretch}{1.3}
  \begin{tabular}{|c||c|c|c|}
  \hline
  $\alpha$ & $r_{_{ISCO}}$ & $L_{_{ISCO}}$ & $E_{_{ISCO}}$ \\\hline\hline
  $0.09$ & $6.252$ & $3.699$ & $0.940$ \\\hline
  $0.2$ & $6.534$ & $3.980$ & $0.937$ \\\hline
  $0.4$ & $6.968$ & $4.470$ & $0.932$ \\
  \hline
  \end{tabular}}
\end{table}

\subsubsection{Radiant energy flux}

In accretion process, the falling particles at infinity from rest will accrete onto the source mass. During the process, the gravitational energy of these falling particles will release and then convert into the electromagnetic radiation \cite{Kato:20008,Page:1974he}. One can express the radiation flux of the accretion disc around the central mass in the following form, which depends on the specific angular momentum, the specific energy, and the angular velocity of the falling test particle \cite{Kato:20008,Page:1974he}
\begin{equation}\label{radflu1}
\mathcal{F}(r)=-\frac{\dot{M}}{4\pi}\frac{\Omega'_{\varphi}}{\sqrt{-g}\left(E-L\Omega_{\varphi}\right)^{2}}\int_{r_{_{ISCO}}}^{r}\left(E-L\Omega_{\varphi}\right)L'dr\,,
\end{equation}
where $\dot{M}$ is the accretion rate and $g=\det(g_{\mu\nu})=-r^{4}\sin^{2}\theta$ is the determinant of the background metric tensor associated with the line element \eqref{le}, so that on the equatorial plane, we have $g=-r^{4}$. Inserting Eqs. \eqref{E2}-\eqref{Omega2} into Eq. \eqref{radflu1} and also, using numerical data in Table \ref{Table1}, one can find an approximate expression for radiation flux as follows
\begin{equation}\label{radflu2}
\begin{split}
\mathcal{F}(r) & \approx-\frac{(1+\alpha)\dot{M}M^{\frac{3}{2}}y_{1}y_{3}\left(\alpha(1+\alpha)M^{2}+r^{2}\right)^{\frac{5}{4}}}{96\pi(r-3M)r^{\frac{3}{2}}
\sqrt{(1+\alpha)My_{2}}\left((1+\alpha)Mr^{3}y_{2}-rf(r) \left(\alpha(1+\alpha)M^{2}+r^{2}\right)^{\frac{7}{2}}\right)^{2}}\\
& \times\left(18(\alpha-2)Mr^{2}+6(\alpha+2)r^{3}-M^{2}\alpha(6M+79r)-4(3+\alpha)\sqrt{3M}r^{\frac{3}{2}}(3M-r)\tanh^{-1}\left[\sqrt{\frac{3M}{r}}\right]\right)\,,
\end{split}
\end{equation}
where we have defined
\begin{equation}\label{y3}
y_{3}\equiv M\alpha r^{3}\left(4\sqrt{\alpha(1+\alpha)M^{2}+r^{2}}+9(1+\alpha)M\right)+4\alpha^{2}(1+\alpha)M^{3}r\left(3(1+\alpha)M-2\sqrt{\alpha(1+\alpha)M^{2}+r^{2}}\right)
-3r^{5}\,.
\end{equation}

\begin{figure}[htb]
\centering
\subfloat[\label{Fig5a} for $\alpha=0.09$ with $r_{_{ISCO}}=6.252$]{\includegraphics[width=0.475\textwidth]{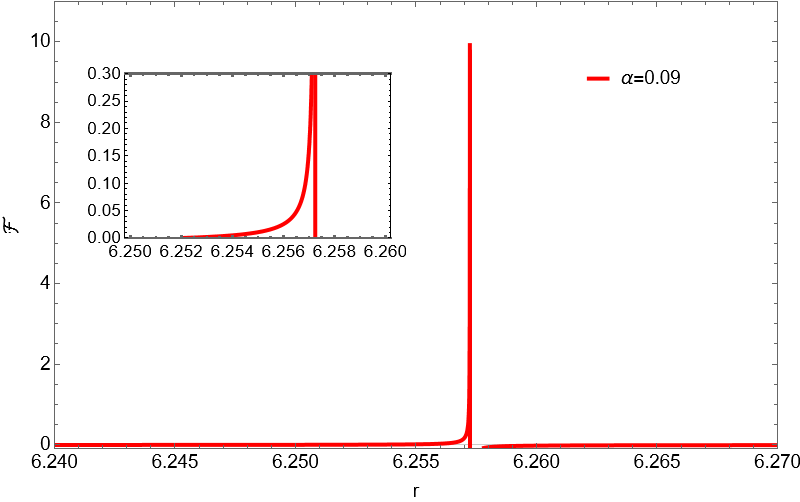}}
\,\,\,
\subfloat[\label{Fig5b} for $\alpha=0.2$ with $r_{_{ISCO}}=6.534$]{\includegraphics[width=0.475\textwidth]{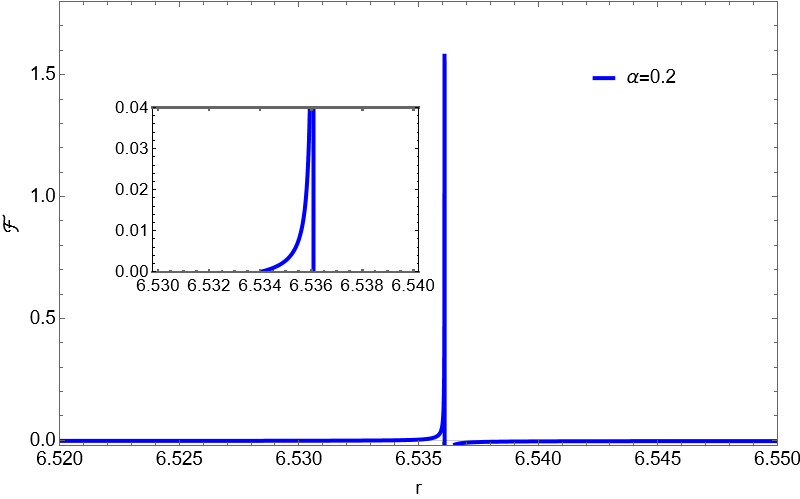}}
\caption{\label{Fig5}\small{\emph{The behavior of $\mathcal{F}(r)$ versus $r$ for the regular static spherically symmetric MOG dark compact object.}}}
\end{figure}

Figure \ref{Fig5} is the illustration of energy flux $\mathcal{F}(r)$ over $r$ associated with the regular MOG dark compact object so that Fig. \ref{Fig5a} is related to $\alpha=0.09$ and Fig. \ref{Fig5b} is related to $\alpha=0.2$. From Fig. \ref{Fig5} we see that the energy flux in the setup is zero at $r<r_{_{ISCO}}$ and then at $r=r_{_{ISCO}}$ it grows from zero to infinity at $r>r_{_{ISCO}}$ and after that, it again becomes zero at far from the source. Comparing Figs. \ref{Fig5a} and \ref{Fig5b} demonstrates that increasing the value of $\alpha$ leads to decrease the energy in the setup.

It also should be noted that the thermodynamical equilibrium is basic demand for the model describing the steady-state accretion disk. As a result, the radiation emitted from the accretion disk surface is equivalently as the black body spectrum \cite{Perez:2017spz,Kato:20008,Page:1974he}. This means that the energy flux and the effective temperature of the accretion disk can be related with the well-known Stefan-Boltzman law $\mathcal{F}(r)=\sigma_{_{SB}}T^{4}$ in which $\sigma_{_{SB}}$ is Stefan-Boltzman constant. Therefore, from this law, one can find the effective temperature $T$ of the accretion disk. Furthermore, at the distance $d$ with the inclination angle $\gamma$ to the central mass, the luminosity of accretion disk can be found as \cite{Perez:2017spz,Kato:20008,Page:1974he}
\begin{equation}\label{luacdi}
L(\upsilon)=4\pi d^{2}I(\upsilon)=\frac{8\cos\gamma}{\pi}\int_{r_{_{ISCO}}}^{r}\int_{0}^{2\pi}\frac{\upsilon_{e}^{3}r}{\exp\left[\frac{\upsilon_{e}}{T}\right]-1}
\,d\varphi\,dr\,,
\end{equation}
where $I(\upsilon)$ is the thermal energy flux as function of frequency $\upsilon$, while $\upsilon_{e}=\upsilon(1+z)$ is the emitted frequency at the redshift $z$. Calculating the luminosity of accretion disk from the above equation is not possible, analytically due to the complexity of the relations.

\subsection{Motion of electrically charged test particle}

Assuming magnetic coupling process \cite{ZNAJEK:1976,Blandford:1977ds,Wang:2002pp,Wang:2003yk,Zahrani:2013up} in the vicinity of regular MOG dark compact object, the energy and angular momentum can be transferred from the dark compact object to the accretion disk. Therefore, on the horizon of the dark compact object, the strength of the magnetic field is expressed as \cite{Hussain:2015cga}
\begin{equation}\label{mgfiho}
B_{h}=\frac{1}{r_{h}}\sqrt{2m_{p}\dot{M}c}\,,
\end{equation}
where the index ($h$) stands for horizon, $c$ is the speed of light, and $m_{p}$ is the magnetization parameter so that $m_{p}=1$ means the equipartition state for the accretion and magnetic coupling process. Theoretical and experimental evidence demonstrate that the magnetic field can be exist in the surroundings of black holes and other compact objects \cite{Koide:2003fj,Hussain:2014cba,Jamil:2014rsa}. Here, we suppose a weak magnetic field whose energy cannot influence the background geometry \cite{Frolov:2011ea}. Accordingly, this type of regular MOG dark compact object is called weakly magnetized.

\begin{figure}[htb]
\centering
\includegraphics[width=0.7\textwidth]{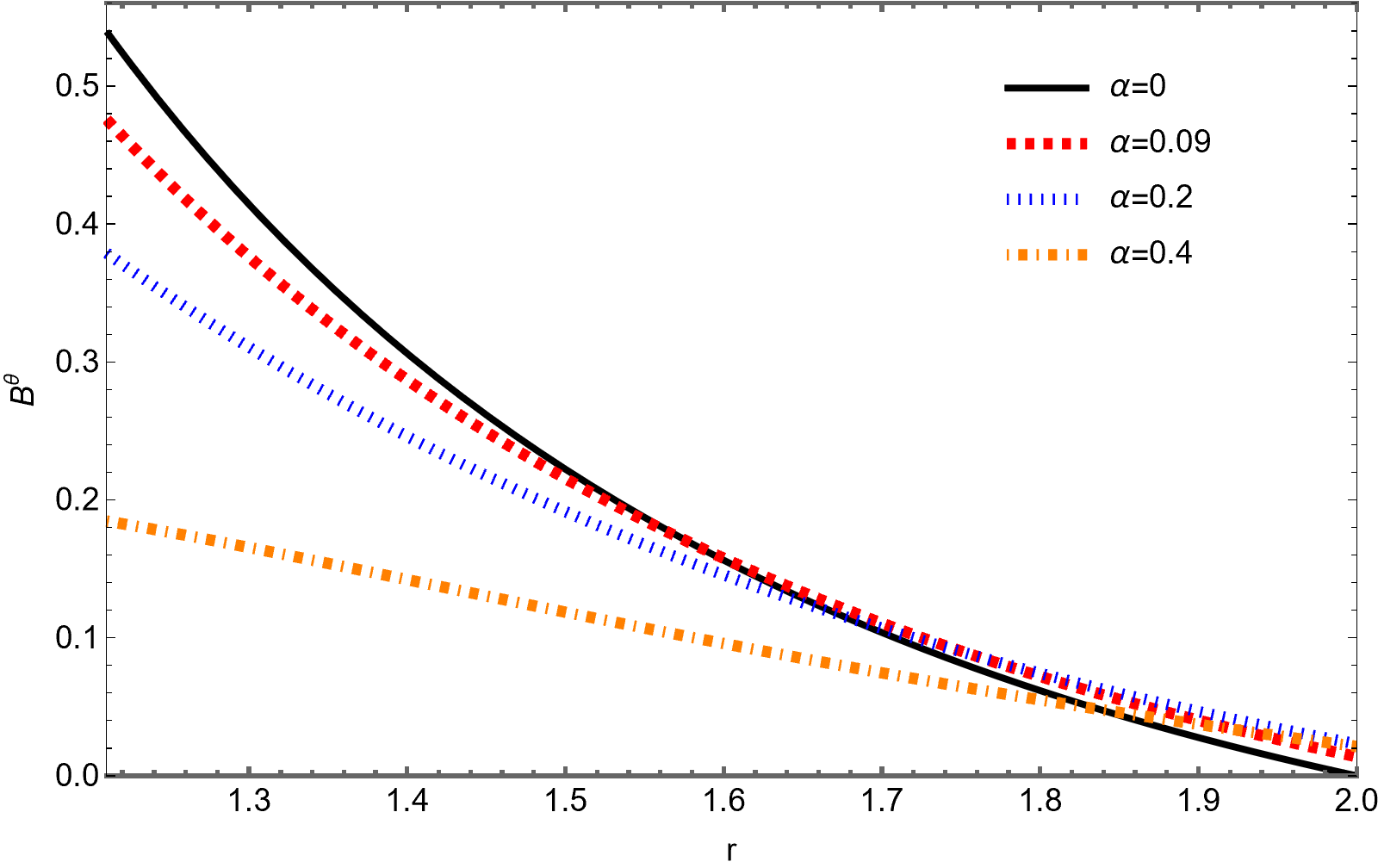}
\caption{\label{Fig6}\small{\emph{The illustration of $B^{\theta}$ around the regular static spherically symmetric MOG dark compact object versus $r$ for different values of $\alpha$, where  we have set $M=1$. The black solid line is for the case of Schwarzschild solution in GR.}}}
\end{figure}

Following the procedure introduced in Refs. \cite{Hussain:2015cga,Zahrani:2013up,Hussain:2014cba}, we aim to calculate the magnetic field in the surroundings of the regular MOG dark compact object. The Killing vectors introduced in Eq. \eqref{killvec} satisfy the following Killing vector equation \cite{Wald:1974np}
\begin{equation}\label{killveceq}
\Box\zeta^{\mu}=0\,,
\end{equation}
where $\Box=\partial_{\mu}\partial^{\mu}$ is the d'Almmbert operator. In Lorentz gauge, the above equation is equivalent with the Maxwell equation for four-potential
\begin{equation}\label{Maeqfp}
A^{\mu}_{;\mu}=0\,,
\end{equation}
in which $A^{\mu}$ is the four-potential and `` ; " shows covariant derivative. The expression
\begin{equation}\label{fopo}
A^{\mu}=\frac{B}{2}\,{}^{(\varphi)}\zeta^{\mu}=\left(0,0,0,\frac{B}{2}\right)\,,
\end{equation}
is related to a weak magnetic field, which is homogeneous at the spatial infinity with the strength $B$. Moreover, the magnetic field four-vector can be defined as follows
\begin{equation}\label{fopo}
B^{\mu}=-\frac{\epsilon^{\mu\nu\lambda\sigma}}{\sqrt{-g}}F_{\lambda\sigma}w_{\nu}\,,
\end{equation}
where $\epsilon^{\mu\nu\lambda\sigma}$ is the Levi-Cività symbol, $F_{\lambda\sigma}=A_{\nu;\mu}-A_{\mu;\nu}$ is the Maxwell tensor, and$w_{\nu}$ is the four-velocity of a local observer at rest, which can be written as
\begin{equation}\label{fovelor}
w_{\nu}=\left(\frac{1}{\sqrt{f(r)}},0,0,0\right)\,.
\end{equation}
Utilizing Eqs. \eqref{Maeqfp}-\eqref{fovelor} results in the magnetic field four-vector expression, which on the equatorial plane is as follows
\begin{equation}\label{mafifve}
B^{\nu}=\left(0,0,-\frac{B\sqrt{f(r)}}{r},0\right)\,.
\end{equation}
It is assumed that the magnetic field is directed upward along the $\mathrm{z}$-axis at spatial infinity \cite{Huang:2014tra}. Figure \ref{Fig6} demonstrates the illustration of $B^{\theta}$ in vicinity of the regular MOG dark compact object versus $r$ for different values of $\alpha$. We see from Fig. \ref{Fig6} that far from the regular MOG dark compact object, the magnetic field is almost vanishing. Also, the effect of $\alpha$ on $B^{\theta}$ is to reduce its strength.

\begin{figure}[htb]
\centering
\includegraphics[width=0.7\textwidth]{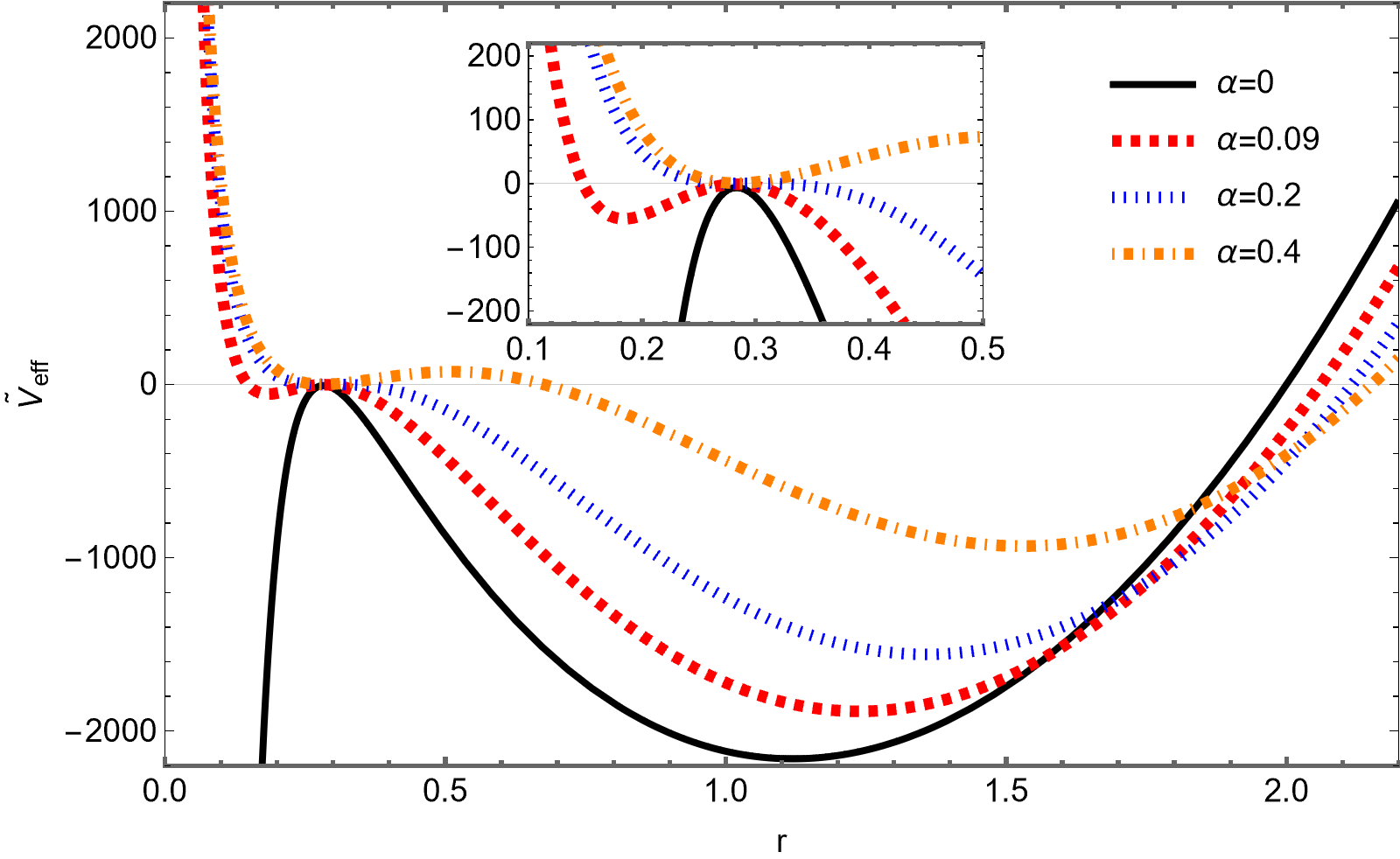}
\caption{\label{Fig7}\small{\emph{The illustration of $\tilde{V}_{eff}$ of the massive electrically charged test particle moving in the spacetime of the regular static spherically symmetric MOG dark compact object versus $r$ for different values of $\alpha$. The black solid line is for the case of Schwarzschild solution in GR.}}}
\end{figure}

The Lagrangian of an electrically charged test particle with rest mass $m$ and electric charge $q$ travelling in the spacetime of the regular MOG dark compact object is expressed as
\begin{equation}\label{lagelch}
\tilde{\mathcal{L}}=\frac{1}{2}g_{\mu\nu}\dot{x}^{\mu}\dot{x}^{\nu}+\frac{1}{m}A_{\mu}\dot{x}^{\mu}\,.
\end{equation}
Similar to the previous section, using Euler-Lagrange equation \eqref{eullag}, on the equatorial plane we can find
\begin{equation}\label{eelch2}
\dot{t}\equiv\tilde{u}^{t}=\frac{\tilde{E}}{f(r)}=\frac{\tilde{E}}{\left(1-\frac{2(1+\alpha)Mr^{2}}{\left(r^{2}+\alpha(1+\alpha)M^{2}\right)^{\frac{3}{2}}}
+\frac{\alpha(1+\alpha)M^{2}r^{2}}{\left(r^{2}+\alpha(1+\alpha)M^{2}\right)^{2}}\right)}\,,
\end{equation}
and
\begin{equation}\label{Lelch2}
\dot{\varphi}\equiv\tilde{u}^{\varphi}=\frac{\tilde{L}}{r^{2}}-\frac{qB}{2m}\,,
\end{equation}
where $\tilde{u}^{\mu}$ is the four-velocity of the electrically charged test particle, $\tilde{E}$ and $\tilde{L}$ are the specific energy and specific angular momentum of the electrically charged test particle respectively. Again, on the equatorial plane, one can employ the normalization condition $\tilde{u}^{\mu}\tilde{u}_{\mu}=1$ to gain
\begin{equation}\label{rdotelch2}
\dot{r}^{2}=\tilde{E}^{2}-\tilde{V}_{eff}\,,
\end{equation}
where $\tilde{V}_{eff}$ is the effective potential of the electrically charged test particle, which is
\begin{equation}\label{efpoelch2}
\begin{split}
\tilde{V}_{eff} & =f(r)\left(1+r^{2}\left(\frac{\tilde{L}}{r^{2}}-\frac{qB}{2m}\right)^{2}\right)\\
& =\left(1-\frac{2(1+\alpha)Mr^{2}}{\left(r^{2}+\alpha(1+\alpha)M^{2}\right)^{\frac{3}{2}}}+\frac{\alpha(1+\alpha)M^{2}r^{2}}{\left(r^{2}+\alpha(1+\alpha)M^{2}\right)^{2}}
\right)\left(1+r^{2}\left(\frac{\tilde{L}}{r^{2}}-\frac{qB}{2m}\right)^{2}\right)\,.
\end{split}
\end{equation}
Figure \ref{Fig7} is the illustration of $\tilde{V}_{eff}$ versus $r$ for the electrically charged test particle moving in the spacetime of the regular MOG dark compact object for different values of $\alpha$. From Fig. \ref{Fig3} we see that increasing $\alpha$ firstly leads to increase the effective potential while in far from the source decreases it.

\subsubsection{Stable circular orbits around regular MOG dark compact object for electrically charged particles}

Similar to previous section, the conditions $E^{2}=V_{eff}$ and $\frac{dV_{eff}}{dr}=0$ must be satisfied for circular orbits. Therefore, one can solve these equations simultaneously by using Eqs. \eqref{fle}, \eqref{eelch2}, and \eqref{Lelch2} to obtain the following relations
\begin{equation}\label{Eelch2}
\tilde{E}^{2}=f(r)\left(1+\frac{r\left(\sqrt{B^{2}q^{2}rf(r)^{2}-m^{2}rf'(r)^{2}+2m^{2}f(r)f'(r)}-Bq\sqrt{r}f(r)\right)^{2}}{m^{2}\left(rf'(r)-2f(r)\right)^{2}}\right)\,,
\end{equation}
\begin{equation}\label{Lelch}
\begin{split}
\tilde{L}=\frac{2r^{\frac{3}{2}}\sqrt{B^{2}q^{2}rf(r)^{2}-m^{2}rf'(r)^{2}+2m^{2}f(r)f'(r)}-Bqr^{3}f'(r)}{4mf(r)-2mrf'(r)}\,,
\end{split}
\end{equation}
\begin{equation}\label{Omegaelch}
\tilde{\Omega}_{\varphi}^{2}=\frac{\left(\sqrt{B^{2}q^{2}rf(r)^{2}-m^{2}rf'(r)^{2}+2m^{2}f(r)f'(r)}-Bq\sqrt{r}f(r)\right)^{2}}{2r\left(f(r)\left(B^{2}q^{2}r^{2}+2m^{2}\right)-r \left(Bq\sqrt{r}\sqrt{B^{2}q^{2}rf(r)^{2}-m^{2}rf'(r)^{2}+2m^{2}f(r)f'(r)}+m^{2}f'(r)\right)\right)}\,.
\end{equation}
Eqs. \eqref{Eelch2}-\eqref{Omegaelch} in the limit of $q\rightarrow 0$ reduce to Eqs. \eqref{E2}-\eqref{Omega2}.

The location of ISCO for the massive electrically charged test particle moving in regular MOG dark compact object satisfies the conditions \eqref{iscocons}. As we previously mentioned, the explicit analytical form of ISCO for the electrically charged test particle associated with the regular MOG dark compact object is not available due to complexity of the metric coefficient \eqref{fle}. Thus, one can numerically solve the equations set \eqref{iscocons} by using, for instance, Wolfram Mathematica (v13.1) to obtain the numerical values of the ISCO for the electrically charged test particle moving in the spacetime of the MOG regular dark compact object. To do this, in Table \ref{Table2} we again set $M=1$ and for different values of $\alpha$ we collect the numerical values of $\tilde{r}_{_{ISCO}}$, $\tilde{L}_{_{ISCO}}$, and $\tilde{E}_{_{ISCO}}$ corresponding with the electrically charged test particle for the regular static spherically symmetric MOG dark compact object. Table \ref{Table2} demonstrates that increasing the value of $\alpha$ causes to increase the ISCO radius of electrically charged test particle associated with the regular MOG dark compact object. Comparing Tables \ref{Table1} and \ref{Table2} shows us that the values of ISCO related to the weakly magnetised regular MOG dark compact object are smaller than the corresponding ones related to the regular MOG dark compact object. Therefore, the electric charge of the test particle and the magnetic field in the vicinity of the source affect the ISCO radius by reducing it.
\begin{table}[htb]
  \centering
  \caption{\label{Table2}\small{\emph{The numerical values of $\tilde{r}_{_{ISCO}}$, $\tilde{L}_{_{ISCO}}$, and $\tilde{E}_{_{ISCO}}$ for an electrically charged test particle moving in the regular static spherically symmetric MOG dark compact object spacetime associated with different values of $\alpha$.}}}
  {\renewcommand{\arraystretch}{1.3}
  \begin{tabular}{|c||c|c|c|}
  \hline
  $\alpha$ & $\tilde{r}_{_{ISCO}}$ & $\tilde{L}_{_{ISCO}}$ & $\tilde{E}_{_{ISCO}}$ \\\hline\hline
  $0.09$ & $0.135$ & $0.844$ & $0.403$ \\\hline
  $0.2$ & $0.303$ & $4.428$ & $0.246$ \\\hline
  $0.4$ & $0.664$ & $21.659$ & $0.134$ \\
  \hline
  \end{tabular}}
\end{table}
Similar to previous section, one can find the energy flux associated with the massive electrically charged particle moving in the regular MOG dark compact object spacetime by inserting Eqs. \eqref{Eelch2}-\eqref{Omegaelch} into Eq. \eqref{radflu1}. Also, the corresponding luminosity of the accretion disk can be found by Eq. \eqref{luacdi}. However, due to the lengthy and complexity of the related equations, it cannot be solved analytically.

\section{Accretion onto regular MOG dark compact object}\label{ARMDCO}

In this section, we aim to find the basic dynamical equations and parameters associated with the accretion onto the regular MOG dark compact object following the procedure performed in Refs. \cite{Nozari:2020swx,Salahshoor:2018plr}. To do this, we take into account the spherically symmetric accretion in the equatorial plane with $\theta=\frac{\pi}{2}$. Additionally, we assume that the accreting matter is inflowing perfect fluid onto the regular MOG dark compact object.

\subsection{Dynamical equations}

The perfect fluid stress-energy tensor is expressed as
\begin{equation}\label{pfset}
T^{\mu\nu}=(p+\rho)v^{\mu}v^{\nu}-pg^{\mu\nu}\,,
\end{equation}
where $p$, $\rho$, and $v^{\mu}$ are pressure, energy density, and four-velocity of the perfect fluid, respectively. On the equatorial plane, the only non–vanishing four-velocity components are $v^{\mu}=(v^{t},v^{r},0,0)$. To be precise, the four-velocity of the perfect fluid $v^{\mu}$ and the four-velocity of the test particle $u^{\mu}$ in previous section are equivalent since the inflowing fluid, in fact, travels on the time-like geodesics creating the accretion disk around the MOG compact object. Therefore, the trajectory of inflowing fluid and the test particle in previous section are identical. In other words, the test particle in the previous section is assumed here as perfect fluid. On the other hand, according to the normalization condition for the four-velocity of the perfect fluid $v^{\mu}v_{\mu}=1$ one can find
\begin{equation}\label{vt}
v^{t}=\frac{\sqrt{f(r)+(v^{r})^{2}}}{f(r)}=\frac{\sqrt{1-\frac{2(1+\alpha)Mr^{2}}{\left(r^{2}+\alpha(1+\alpha)M^{2}\right)^{\frac{3}{2}}}+\frac{\alpha(1+\alpha)M^{2}r^{2}}
{\left(r^{2}+\alpha(1+\alpha)M^{2}\right)^{2}}+(v^{r})^{2}}}{1-\frac{2(1+\alpha)Mr^{2}}{\left(r^{2}+\alpha(1+\alpha)M^{2}\right)^{\frac{3}{2}}}+\frac{\alpha(1+\alpha)
M^{2}r^{2}}{\left(r^{2}+\alpha(1+\alpha)M^{2}\right)^{2}}}\,,
\end{equation}
where the condition $v^{r}<0$ must be satisfied since the accretion is an inward flow of matter while the assumption $v^{t}>0$ is taken into account because we interested in forward flow in time.

From the conservation of the stress-energy tensor, i.e., $T^{\mu\nu}_{;\nu}=0$ in which ($;$) stands for covariant derivative, we can find the following relation
\begin{equation}\label{conset}
(p+\rho)v^{r}r^{2}\sqrt{1-\frac{2(1+\alpha)Mr^{2}}{\left(r^{2}+\alpha(1+\alpha)M^{2}\right)^{\frac{3}{2}}}+\frac{\alpha(1+\alpha)M^{2}r^{2}}
{\left(r^{2}+\alpha(1+\alpha)M^{2}\right)^{2}}+(v^{r})^{2}}=A_{0}\,,
\end{equation}
where $A_{0}$ is a constant of integration. Additionally, we can project the stress-energy tensor conservation law onto the perfect fluid four-velocity to the form of
\begin{equation}\label{proj1}
v_{\mu}T^{\mu\nu}_{;\nu}=0\,,
\end{equation}
which results in the following relation
\begin{equation}\label{proj2}
\frac{\rho'}{p+\rho}+\frac{(v^{r})'}{v^{r}}+\frac{2}{r}=0\,.
\end{equation}
By integrating, the last equation yields
\begin{equation}\label{proj3}
r^{2}v^{r}\exp{\left[\int\frac{d\rho}{p+\rho}\right]}=-A_{1}\,,
\end{equation}
where $A_{1}$ is a constant of integration. Since the condition $u^{r}<0$ holds, one can deduce
\begin{equation}\label{proj4}
(p+\rho)\exp{\left[-\int\frac{d\rho}{p+\rho}\right]}\sqrt{1-\frac{2(1+\alpha)Mr^{2}}{\left(r^{2}+\alpha(1+\alpha)M^{2}\right)^{\frac{3}{2}}}+\frac{\alpha(1+\alpha)
M^{2}r^{2}}{\left(r^{2}+\alpha(1+\alpha)M^{2}\right)^{2}}+(v^{r})^{2}}=A_{2}\,,
\end{equation}
where $A_{2}$ is an integration constant.

Equation of mass flux in the setup is given by
\begin{equation}\label{emf1}
(\rho v^{\mu})_{;\mu}=0\,,
\end{equation}
where on the equatorial plane results in the following relation
\begin{equation}\label{emf2}
\rho v^{r}r^{2}=A_{3}\,,
\end{equation}
where $A_{3}$ is an integration constant.

\subsection{Dynamical parameters}

Isothermal fluids with the equation of state $p=\omega\rho$ where $\omega$ is the equation of state parameter are taken into account. During the motion of these fluids, the temperature remains constant. Consequently, Eqs. \eqref{proj3}, \eqref{proj4}, and \eqref{emf2} yields
\begin{equation}\label{a4}
\frac{p+\rho}{\rho}\sqrt{1-\frac{2(1+\alpha)Mr^{2}}{\left(r^{2}+\alpha(1+\alpha)M^{2}\right)^{\frac{3}{2}}}+\frac{\alpha(1+\alpha)
M^{2}r^{2}}{\left(r^{2}+\alpha(1+\alpha)M^{2}\right)^{2}}+(v^{r})^{2}}=A_{4}\,,
\end{equation}
where $A_{4}$ is an integration constant. Inserting $p=\omega\rho$ into the lest equation yields the $v^{r}$ as follows
\begin{equation}\label{vr}
v^{r}=\left(\frac{1}{\omega+1}\right)\sqrt{A_{4}^{2}-\left(\omega+1\right)^{2}\left(1-\frac{2(1+\alpha)Mr^{2}}{\left(r^{2}+\alpha(1+\alpha)M^{2}\right)^{\frac{3}{2}}}
+\frac{\alpha(1+\alpha)M^{2}r^{2}}{\left(r^{2}+\alpha(1+\alpha)M^{2}\right)^{2}}\right)}\,.
\end{equation}

\begin{figure}[htb]
\centering
\includegraphics[width=0.7\textwidth]{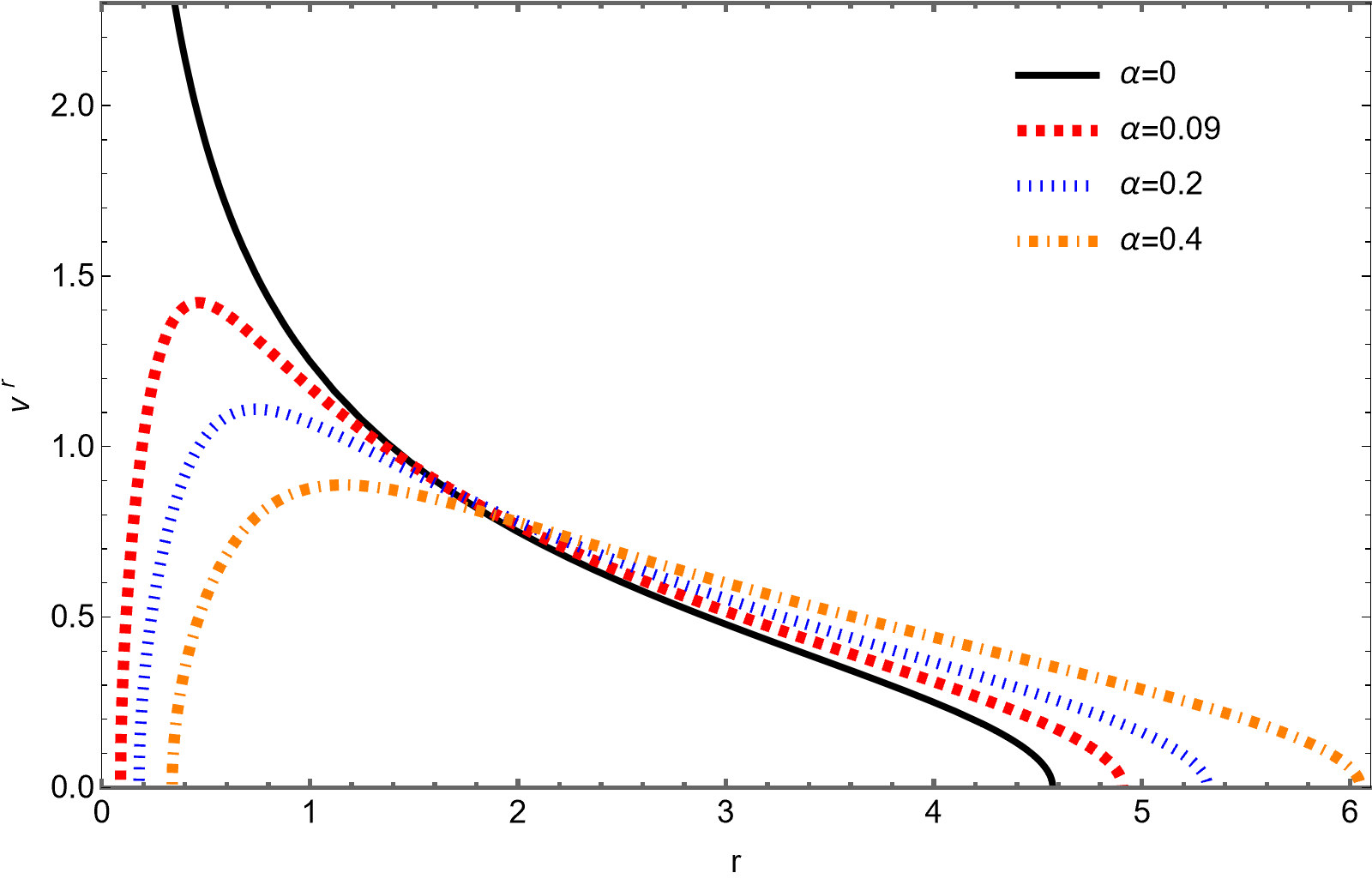}
\caption{\label{Fig11}\small{\emph{The illustration of $v^{r}$ of the regular static spherically symmetric MOG dark compact object versus $r$ for different values of $\alpha$. The black solid line is for the case of Schwarzschild solution in GR.}}}
\end{figure}

Figure \ref{Fig11} is the graph of $v^{r}$ versus $r$ for the regular static spherically symmetric MOG dark compact object in comparison with Schwarzschild black hole in GR. From Fig. \ref{Fig11}, we see that the fluid with the radial element of its four-velocity corresponding with each curve (associated with each value of $\alpha$) begins to move towards the regular MOG dark compact object from rest at large $r$, as previously mentioned. Then, it approaches the regular static spherically symmetric MOG dark compact object to again reach the rest state. Furthermore, from Fig. \ref{Fig4}, we see that decreasing the parameter $\alpha$ leads to increase the value of $v^{r}$ far from the source. Moreover, the curve of Schwarzschild case goes to infinity by approaching the regular static spherically symmetric MOG dark compact object.

\begin{figure}[htb]
\centering
\includegraphics[width=0.7\textwidth]{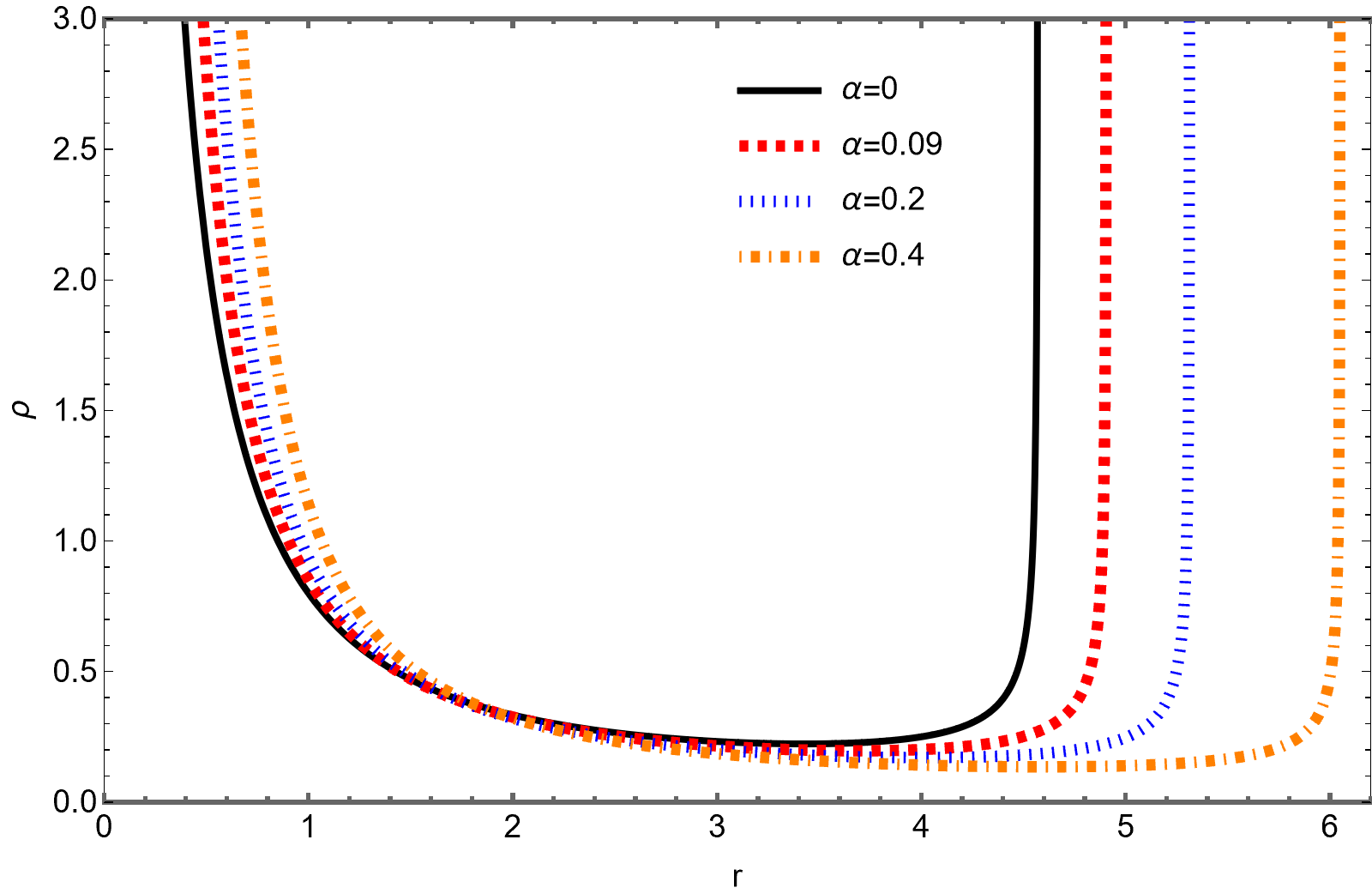}
\caption{\label{Fig12}\small{\emph{The illustration of $\rho$ of the regular static spherically symmetric MOG dark compact object versus $r$ for different values of $\alpha$. The black solid line is for the case of Schwarzschild solution in GR.}}}
\end{figure}

The proper energy density of the fluid can easily be determined as follows
\begin{equation}\label{rho}
\rho=\left(\frac{A_{3}}{r^{2}}\right)\frac{(\omega+1)}{\sqrt{A_{4}^{2}-\left(\omega+1\right)^{2}\left(1-\frac{2(1+\alpha)Mr^{2}}{\left(r^{2}+\alpha(1+\alpha)M^{2}
\right)^{\frac{3}{2}}}+\frac{\alpha(1+\alpha)M^{2}r^{2}}{\left(r^{2}+\alpha(1+\alpha)M^{2}\right)^{2}}\right)}}\,.
\end{equation}
Figure \ref{Fig12} demonstrates the illustration of $\rho$ versus $r$ for the regular static spherically symmetric MOG dark compact object in comparison with Schwarzschild black hole in GR.

\subsection{Mass evolution}

The central source mass of a black holes as well as a dark compact object is a dynamic quantity over time. Accretion process, for example, leads to grow their mass by accreting the surrounding matter onto them. The mass change measure or accretion rate of the regular MOG dark compact object can be obtained through $\dot{M}\equiv\frac{dM}{dt}=-\int T^{r}_{t}dS$ in which the surface element of the object is $dS=\left(\sqrt{-g}\right)d\theta d\varphi$ and also $T^{r}_{t}=(p+\rho)v_{t}v^{r}$. Consequently, the accretion rate $\dot{M}$ can be obtained as
\begin{equation}\label{mdot1}
\dot{M}=-4\pi r^{2}v^{r}(p+\rho)\sqrt{1-\frac{2(1+\alpha)Mr^{2}}{\left(r^{2}+\alpha(1+\alpha)M^{2}\right)^{\frac{3}{2}}}+\frac{\alpha(1+\alpha)
M^{2}r^{2}}{\left(r^{2}+\alpha(1+\alpha)M^{2}\right)^{2}}+(v^{r})^{2}}\equiv-4\pi A_{0}\,,
\end{equation}
where the definitions $A_{0}\equiv-A_{1}A_{2}$ and $A_{2}\equiv\left(p_{\infty}+\rho_{\infty}\right)\sqrt{f\left(r_{\infty}\right)}$ are assumed. Finally, we gain
\begin{equation}\label{mdot2}
\dot{M}=4\pi A_{1}M^{2}\left(p_{\infty}+\rho_{\infty}\right)\sqrt{f\left(r_{\infty}\right)}\,.
\end{equation}
One can use Eq. \eqref{mdot2} to obtain a relation between the initial mass $M_{i}$ and the mass in arbitrary time $t$ as follows
\begin{equation}\label{a18}
M_{t}=\frac{M_{i}}{1-\frac{t}{t_{cr}}}\,,
\end{equation}
where the critical accretion time is defined as $t_{cr}=\left(4\pi A_{1}M_{i}(p+\rho)\sqrt{f\left(r_{\infty}\right)}\right)^{-1}$. At $t=t_{cr}$, the mass of the regular MOG dark compact object approaches infinity in a finite time.

\subsection{Critical Accretion}

In accretion process, the inward flow of the fluid from rest at far from the source (regular MOG dark compact object) begins to move and continues to accelerate due to the gravitational field of the central source. During the inward flow motion of the fluid towards the source, it reaches sonic (critical) point, where the four-velocity of the fluid coincides the local speed of sound $c_{s}$. From this critical point to the central source, the inward flow accelerated motion has supersonic velocities. A radial velocity gradient is needed to find the critical point.

The derivatives of Eqs. \eqref{emf2} and \eqref{a4} yield
\begin{equation}\label{crit1}
\frac{\rho'}{\rho}+\frac{(v^{r})'}{(v^{r})}+\frac{2}{r}=0\,,
\end{equation}
and
\begin{equation}\label{crit2}
\begin{split}
& \frac{\rho'}{\rho}\left(\frac{d\ln[p+\rho]}{d\ln[\rho]}-1\right)+\frac{v^{r}(v^{r})'\left(\alpha(1+\alpha)M^2+r^2\right)^{\frac{7}{2}}+y_{2}
(1+\alpha)Mr}{\left(\alpha(1+\alpha)M^2+r^2\right)^{\frac{7}{2}}\left(1-\frac{2(1+\alpha)Mr^{2}}{\left(r^{2}+\alpha(1+\alpha)M^{2}\right)
^{\frac{3}{2}}}+\frac{\alpha(1+\alpha)M^{2}r^{2}}{\left(r^{2}+\alpha(1+\alpha)M^{2}\right)^{2}}+(v^{r})^{2}\right)}=0\,.
\end{split}
\end{equation}
Eqs. \eqref{crit1} and \eqref{crit2} result in the following relation
\begin{equation}\label{crit3}
\frac{d\ln[v^{r}]}{d\ln[r]}=\frac{\mathcal{D}_{1}}{\mathcal{D}_{2}}\,,
\end{equation}
where we defined
\begin{equation}\label{D1}
\mathcal{D}_{1}\equiv-2V^{2}+\frac{y_{2}(1+\alpha)Mr^{2}}{\left(\alpha(1+\alpha)M^2+r^2\right)^{\frac{7}{2}}\left(1-\frac{2(1+\alpha)Mr^{2}}{\left(r^{2}+\alpha(1+\alpha)
M^{2}\right)^{\frac{3}{2}}}+\frac{\alpha(1+\alpha)M^{2}r^{2}}{\left(r^{2}+\alpha(1+\alpha)M^{2}\right)^{2}}+(v^{r})^{2}\right)}\,,
\end{equation}
and
\begin{equation}\label{D2}
\mathcal{D}_{2}\equiv V^{2}-\frac{(v^{r})^{2}}{1-\frac{2(1+\alpha)Mr^{2}}{\left(r^{2}+\alpha(1+\alpha)M^{2}\right)^{\frac{3}{2}}}+\frac{\alpha(1+\alpha)
M^{2}r^{2}}{\left(r^{2}+\alpha(1+\alpha)M^{2}\right)^{2}}+(v^{r})^{2}}\,,
\end{equation}
in which
\begin{equation}\label{V2}
V^{2}\equiv\frac{d\ln[p+\rho]}{d\ln[\rho]}-1\,.
\end{equation}
When the condition $\mathcal{D}_{1}=\mathcal{D}_{2}=0$ is satisfied, the critical points occur. This condition first gives us
\begin{equation}\label{Vcr2}
V_{cr}^{2}=\frac{rf'(r)}{4f(r)+rf'(r)}\,,
\end{equation}
so that the positivity of its denominator determines the range of the critical radius by the following inequality
\begin{equation}\label{racrra}
\begin{split}
& 4\left(1-\frac{2(1+\alpha)Mr^{2}}{\left(r^{2}+\alpha(1+\alpha)M^{2}\right)^{\frac{3}{2}}}+\frac{\alpha(1+\alpha)M^{2}r^{2}}{\left(r^{2}+\alpha(1+\alpha)M^{2}\right)^{2}}
\right)+\frac{2y_{2}(1+\alpha)Mr^{2}}{\left(\alpha(1+\alpha)M^2+r^2\right)^{\frac{7}{2}}}>0\,.
\end{split}
\end{equation}
Additionally, the condition for critical points give us
\begin{equation}\label{vcrr2}
\begin{split}
(v^{r}_{cr})^{2} & =\frac{1}{4}rf'(r)=\frac{y_{2}(1+\alpha)Mr^{2}}{2\left(\alpha(1+\alpha)M^{2}+r^{2}\right)^{\frac{7}{2}}}\,,
\end{split}
\end{equation}
where the index ($cr$) in Eqs. \eqref{Vcr2} and \eqref{vcrr2} stands for critical values. Finally, the local sound speed $c_{s}^{2}=\frac{dp}{d\rho}$ can be found as
\begin{equation}\label{lcs}
c_{s}^{2}=-1+A_{4}\sqrt{1-\frac{2(1+\alpha)Mr^{2}}{\left(r^{2}+\alpha(1+\alpha)M^{2}\right)^{\frac{3}{2}}}+\frac{\alpha(1+\alpha)
M^{2}r^{2}}{\left(r^{2}+\alpha(1+\alpha)M^{2}\right)^{2}}+(v^{r})^{2}}\,.
\end{equation}

\section{Summary and Conclusions}\label{SaC}

In this paper, we explored the accretion onto the regular spherically symmetric MOG dark compact object as well the electrically neutral and charged particles motion in its spacetime, by following the Lagrangian formalism. We found out that the effective potential of the neutral particle moving in the spacetime of the regular MOG dark compact object increases by increasing the value of the parameter $\alpha$, while for the effective potential of the electrically charged particle, it is not the case far from the source. Moreover, we demonstrated that the parameter $\alpha$ of the MOG setup amplifies the specific energy and angular momentum of the test particle, while it decreases the angular velocity. We also showed that the parameter $\alpha$ increases the ISCO radius of either electrically neutral or charged test particles. We saw, however, that the ISCO radius of the electrically neutral (charged) particle associated with the regular MOG dark compact object is larger (smaller) than the corresponding one in the Schwarzschild black hole in GR. By treating the energy flux of the accretion disk related to the neutral particle, we proved that the energy flux peaks after reaching the ISCO and then falls to zero extremely fast, while the parameter $\alpha$ of the MOG setup decreases it. Furthermore, the radial component of the four-velocity and the energy density of the accreting fluid reduce by growing the parameter $\alpha$ near the source, while it is not the case at far from the source.

\end{document}